\documentclass[journal]{IEEEtran}

\usepackage{cite}
\usepackage{graphicx}
\usepackage{subfigure}
\usepackage{amssymb,amsmath} 
\usepackage{multirow}

\begin{document}

\title{Physics Methods for the Simulation of Photoionisation}

\author{Tullio Basaglia, Matej Bati\v{c}, Min Cheol Han, Gabriela Hoff, Chan Hyeong Kim, Han Sung Kim, Maria Grazia Pia and Paolo Saracco% <-this % stops a space
\thanks{Manuscript received 15 November 2013.}% <-this % stops a space
\thanks{This work has been partly funded by CNPq BEX6460/10-0 grant, Brazil.}% <-this % stops a space
\thanks{T. Basaglia is with CERN, CH-1211, Geneva, Switzerland (e-mail: Tullio.Basaglia@cern.ch).}
\thanks{M. Bati\v{c} was with INFN Sezione di Genova, Genova, Italy 
             (e-mail: Batic.Matej@gmail.com); he is now with  
             Sinergise, 1000 Ljubljana, Slovenia.}
\thanks{M. C. Han, C. H. Kim and H. S. Kim are with are with 
	the Department of Nuclear Engineering, Hanyang University, 
        Seoul 133-791, Korea 
	(e-mail: mchan@hanyang.ac.kr, chkim@hanyang.ac.kr, hsungman@naver.com).}
\thanks{G. Hoff is with  
             Pontificia Universidade Catolica do Rio Grande do Sul, Brazil (e-mail:ghoff.gesic@gmail.com).}
\thanks{M. G. Pia and P. Saracco are  with INFN Sezione di Genova, Via Dodecaneso 33, I-16146 Genova, Italy 
	(phone: +39 010 3536328, fax: +39 010 313358, e-mail:
	MariaGrazia.Pia@ge.infn.it, Paolo.Saracco@ge.infn.it).}
}

\maketitle

\pagestyle{empty}
\thispagestyle{empty}

\begin{abstract}
Several physics methods for the simulation of the photoelectric effect are
quantitatively evaluated with respect to a large collection of experimental data 
retrieved from the literature.
They include theoretical and empirical calculations of total and partial cross
sections, and calculations of the photoelectron angular distribution.
Some of these models are currently implemented in general purpose Monte Carlo
systems; some have been implemented and evaluated  for possible use in Monte Carlo particle transport for the first time in this
study.

\end{abstract}
%\begin{keywords}
%Monte Carlo, simulation, Geant4, X-rays
%\end{keywords}

% -----------------------------------------------------------------------------------------

\section{Introduction}
\label{sec_intro}
\PARstart{P}{hotoionization} is important in various experimental
domains, such as material analysis applications, astrophysics, photon science 
and bio-medical physics.
As one of the interactions photons undergo in matter, it is relevant
in experimental methods concerned with the energy deposition
resulting from photons as primary or secondary particles.
Apart from elastic scattering at very low energies, photoionization is the
dominant photon interaction in the low energy r\'egime: as an example, below
approximately 100~keV for target materials of atomic number around 30, and below
approximately 700~keV for heavy target materials of atomic number
close to 90.
Photoionization is also experimentally relevant for the secondary atomic processes that it
induces, X-ray fluorescence and Auger electron emission, which are play a relevant role
in many physics research contexts and technological applications.
Extensive reviews, that cover both the theoretical and experimental aspects of
this process, can be found in the literature, for instance in
\cite{pratt_1973,pratt_1973_err,samson_1976,kelly_1990,starace_2006,amusia_1990,berkowitz_2002}
(this list of references is not intended to be exhaustive).

This paper is concerned with modeling the physics of photoionization under a
pragmatic perspective: the simulation of this process in general purpose Monte
Carlo codes for particle transport.
%Photon interactions with matter play a critical role in these systems \cite{hubbell_2006};
%their modeling presents some peculiarities, because the software must satisfy
%concurrent requirements of physical accuracy and computational performance.

%Various compilations of photoionization cross sections \cite{scofield_1973,chantler_1995,chantler_2000,epdl97,} are available,
%based on theoretical calculations or on experimental data.
%In addition empirical analytical formulations \cite{biggs1,biggs2,biggs3,ebel} 
%are 

Calculations for the simulation of the photoelectric effect are implemented in
all general purpose Monte Carlo systems, nevertheless a comprehensive,
quantitative appraisal of their validity is not yet documented in the
literature.
Assessments reported in the literature usually concern comparisons of cross
sections with NIST reference values, such as \cite{tns_nist}, or involve complex
observables resulting from many physics processes in the full simulation of an
experimental set-up, such as \cite{chica_2009}.
In this respect, it is worthwhile to note that the validation of simulation
models implies their comparison with experimental measurements
\cite{trucano_what}: comparisons with tabulations of theoretical calculations or
analytical parameterizations, such as those that are reported in
\cite{cirrone2010} as validation of Geant4 \cite{g4nim,g4tns} photon interaction cross
sections, do not constitute a validation of the simulation software.
Some relatively recent theoretical calculations and empirical analytical
formulations documented in the literature have not been yet exploited in large
scale Monte Carlo codes, nor have been comparatively evaluated in terms of
accuracy and computational requirements with respect to currently used
simulation methods.

%This paper evaluates the calculation methods adopted by general-purpose Monte
%Carlo systems for the simulation of the photoelectric effect and other modeling
%approaches not yet implemented in these codes, to identify the state of the art
%for the simulation of this process.
%The accuracy of cross section and angular distribution calculations for the
%simulation of the photoelectric effect 
The accuracy of simulation methods is quantified through statistical comparison with a wide
collection of experimental data retrieved from the literature.
%the evaluation of their physics capabilities is complemented by the estimate of their
%computational performance.
These results provide guidance for the selection of physics models in simulation
applications in response to the requirements of physics accuracy and
computational speed pertinent to different experimental scenarios.

Special emphasis is devoted to the validation and possible improvement of
photoionization simulation in Geant4; nevertheless, the results documented in
this paper provide information relevant to other Monte Carlo systems as well.

The simulation of the atomic relaxation following the ionization of an atom has
been treated in previous publications \cite{tns_relax, tns_relax_nist,
tns_relax_prob, tns_binding}, threfore it is not included in the scope of this
paper.

% -----------------------------------------------------------------------------------------

\section{Physics overview}
\label{sec_physics}

Photoionization has been the object of theoretical and experimental interest for
several decades; only a brief overview is included here to facilitate the
comprehension of the software features and the simulation validation results
documented in this paper.

In the photoelectric effect a photon disappears and an electron is ejected from
an atom.
The energy of the photoelectron corresponds to the difference between the energy
of the absorbed photon and the energy binding the electron to the atom.
%The K-shell electrons are the most tightly bound, and are the most important
%contributors to the atomic photoeffect cross section in most cases.

%This paper deals with the simulation of the photoelectric effect within the
%scope of general purpose Monte Carlo codes for particle transport.

%The environment can have significant effect on the cross sections near the
%photoionization thresholds of both inner and outer shell electrons; due to the
%limitations of their underlying physics assumptions, current general purpose
%Monte Carlo codes codes are not usually exploited for the simulation of
%esperimental scenarios involving EXAFS (Extended X-ray Absorption Fine
%Structure), XANES (X-ray Absorption Near Edge Structure) and other techniques
%for which detailed accounting of material structure is required.

The study reported here focuses on the evaluation and validation of basic physics
features relevant to the simulation of the photoelectric effect: atomic
cross sections and photoelectron angular distributions.
%Computational algorithms pertaining to the how these basic modeling entities are
%used in the transport environment, such as methods for dealing with the
%macroscopic cross sections for compounds or mixtures
%\cite{egs_compound1,egs_compound2}, are not discussed here.

\subsection{Total and partial cross sections}

The photoelectric cross section as a function of energy exhibits a
characteristic sawtooth behavior corresponding to absorption edges, as the
binding energy of each electron subshell is attained and corresponding
photoionization is allowed to occur.
%An example of photoelectric cross section is shown in Fig. \ref{fig_photoel}.

Early theoretical calculations of photoionization cross sections were
limited to the K shell; they are tipified by the papers of Pratt
\cite{pratt_1960}, providing the asymptotic behavior for arbitrarily high
energies, and Pratt et al. \cite{pratt_1964}, reporting calculations in the
energy range between 200~keV and 2~MeV.
Only at a later stage more extensive calculations became available: 
Rakavy and Ron \cite{rakavy_1967} calculated cross sections for all subshells of
five elements over the energy range 1~keV to 2~MeV,
Schmickley and Pratt \cite{schmickley_1967} reported cross sections for K to M
shells for three elements from 412 to 1332~keV.

Scofield's non-relativistic calculations \cite{scofield_1973} in a
Hartree-Slater framework represented a major advancement in the field, as they
covered systematically all subshells over the whole periodic table of the
elements.
More recent calculations were performed by Chantler
\cite{chantler_1995,chantler_2000} in a self-consistent relativistic
Dirac-Hartree-Fock framework.

Various empirical formulations of photoionization cross sections are reported in the
literature, such as in 
\cite{hubbell_1969, biggs3, ebel_2003}.
They derive from fits to experimental data, parameterizations of theoretical
calculations and semi-empirical methods involving both measured data and
theoretical considerations.

Computational performance imposes constraints on the complexity of physics
calculations to be performed in the course of simulation: hence the analysis in
this paper is limited to theoretical cross sections for which tabulations of
pre-calculated values are available and to empirical models that are
expressed by means of simple analytical formulations.
To be relevant for general purpose Monte Carlo systems, tabulated data should
cover the whole periodic table of elements and an extended energy range.

The photoelectric cross section compilations considered in this study are
summarized in Table~\ref{tab_compilations}.

% Table generated by Excel2LaTeX from sheet 'Coverage'
\begin{table}[htbp]
  \centering
  \caption{Compilations of photoionization cross sections}
    \begin{tabular}{lrrrrc}
   \hline
    \textbf{Compilation} & \multicolumn{2}{c}{\textbf{Energy range}} & \multicolumn{2}{c}{\textbf{Z range}} & \textbf{Shell} \\
    \hline
    Biggs and Lighthill 	\cite{biggs3}				&  10 eV 	& 100 GeV  &  1	& 100 	&  \\
    Brennan and Cowan \cite{brennan_cowan_1992} 		&  30 eV     & 700 keV 	&  3	& 92 		&  \\
    Chantler \cite{chantler_1995,chantler_2000}			&  10 eV 	& 433 keV 	&  1	& 92 		& K \\
    Ebel \cite{ebel_2003}  								&  1 keV 	&  300 keV 	&  1	& 92 		& all \\
    Elam  \cite{elam_2002} 							&  100 eV 	& 1 MeV 	&  1	& 98 		&  \\
    EPDL97 	\cite{epdl97}							&  10 eV 	& 100 GeV  &  1	& 100 	& all \\
    Henke \cite{henke_1982, henke_1993} 				&  10 eV 	&  30 keV 	&  1	& 92 		&  \\
    McMaster \cite{mcmaster_1969,shaltout_2006} 		&  1 keV 	&  700 keV 	&  1	& 94 		&  \\
    PHOTX \cite{photx} 							&  1 keV 	& 100 MeV 	&  1	& 100 	&  \\
    RTAB  \cite{rtab} 								&  10 eV 	&  30 keV 	&  1	& 99 		& all \\
    Scofield \cite{scofield_1973}						&  1 keV 	& 1.5 MeV 	&  1	& 100 	& all \\
    Storm and Israel \cite{storm_israel_photon} 			&  1 keV 	& 100 MeV 	&  1	& 100 	&  \\
    Veigele 	\cite{veigele_1973} 						&  100 eV 	& 1 MeV 	&  1	& 94 		&  \\
    XCOM  \cite{xcom} 							&  1 keV 	& 500 keV 	&  1	& 100 	&  \\
    \hline
    \end{tabular}%
  \label{tab_compilations}%
\end{table}%

\subsection{Angular distribution}

%\begin{figure}
%\centerline{\includegraphics[angle=0,width=8.5cm]{dummy}}
%\caption{Cross section for the photoionization of ?? as a function of photon energy.}
%\label{fig_photoel}
%\end{figure}

Fischer’s non-relativistic theory \cite{fischer_1931} was derived for use in the
low energy region.
The first relativistic treatment of the photoelectric effect was given by Sauter
\cite{sauter_1931,sauter_1931a}, who calculated the K-shell cross section in the Born
approximation; it is valid to the lowest order in Z$\alpha/\beta$ (where
Z is the atomic number of the target, $\alpha$ is the fine structure constant and $\beta$ is $v/c$).
%A comparison of these theories is discussed in \cite{davisson_evans_1952},
%which showed that Sauter's theory applies even in the non-relativistic realm, 
%despite being derived for relativistic electrons.
Gavrila \cite{gavrila_1959} and Nagel \cite{nagel_1963} extended Sauter's results
to the next order in Z$\alpha/\beta$.
Further calculations by Gavrila are available for the L shel \cite{gavrila_1961}.

%% -----------------------------------------------------------------------------------------

\section{Photoionization in Monte Carlo codes}
\label{sec_mc}

General purpose Monte Carlo  codes consider single photon interactions with
isolated atoms in their ground state; they neglect interactions with ions and
excited states, and multiple ionizations.
Photon interactions are treated regardless of the environment of the target
medium: this assumption neglects solid state effects and other features related to
the molecular structure of the medium.

The original version of EGS4 \cite{egs4} calculated photoelectric total cross
sections based on Storm and Israel's tables \cite{storm_israel_photon} and
generated the photoelectron with the same direction as the incident photon.
Later evolutions introduced the use of PHOTX \cite{photx} cross sections
\cite{sakamoto} and the generation of the photoelectron angular distribution
\cite{egs4_angular} based on Sauter's theory \cite{sauter_1931}.
These features are currently implemented in EGS5 \cite{egs5}.
EGSnrc \cite{egsnrc} provides the option of calculating total photoelectric
cross sections based on Storm and Israel's tables as originally in EGS4 or on a
fit to XCOM \cite{xcom} cross sections, while it uses subshell cross sections
based on EPDL \cite{epdl97}.
It samples the photoelectron angular distribution according to the method
described in \cite{egs4_angular} based on Sauter's theory.

ETRAN \cite{etran} uses Scofield's 1973 \cite{scofield_1973} cross sections for
energies from 1 keV to 1.5 MeV and extends them to higher energies by exploiting
Hubbell's method \cite{hubbell_1969} to connect the values at 1.5 MeV to the
asymptotic high energy limit calculated by Pratt \cite{pratt_1960}.
It samples the direction of the photoelectron from Fischer's
\cite{fischer_1931} distribution for electron energies below 50~keV and from the
Sauter \cite{sauter_1931} distribution for higher energies.

FLUKA \cite{fluka1,fluka2} calculates photoelectric cross sections based on
EPDL97 and samples the photoelectron direction according to Sauter's theory
\cite{sauter_1931}.

ITS \cite{its5} calculates photoelectric cross sections based on 
Scofield's 1973 non-renormalized values.
The angle of the photoelectron with respect to the parent photon is described by
Fischer's distribution \cite{fischer_1931} at lower energies and by Sauter's
\cite{sauter_1931} formula at higher energies.

MCNP5 \cite{mcnp5} and MCNPX \cite{mcnpx27e} provide different options of data
libraries for the calculation of photoelectric cross sections: two version of
EPDL (EPDL97 \cite{epdl97} and EPDL89 \cite{epdl89}), and ENDF/B-IV
\cite{ENDFB-IV} data complemented by Storm and Israel's tables
\cite{storm_israel_photon} for atomic numbers greater than 83.

In the first version of Penelope including photon transport \cite{sempau_1997}
photoelectric cross sections were interpolated from XCOM; in more recent
versions \cite{penelope2008,penelope2011} they are interpolated
from EPDL97 tabulations.
The photoelectron angular distribution is sampled from Sauter's differential
cross section for the K shell \cite{sauter_1931}.

GEANT 3 \cite{geant3} calculated total photoionization cross sections based on 
Biggs and Lighthill's \cite{biggs3} parameterizations; the probability of ionization of the K shell and L 
subshells was estimated by parameterizations of the jump ratios deriving from 
Veigele's \cite{veigele_1973} tables.
The angular distribution of the photoelectron was sampled for the K shell and
for the L$_1$, L$_2$ and L$_3$ subshells based on Sauter's
\cite{sauter_1931,sauter_1931a} and Gavrila's \cite{gavrila_1959,gavrila_1961}
calculations.

The Geant4 toolkit encompasses various implementations of the photoelectric
effect.
The overview summarized here concerns the latest version at the time of the 2013 IEEE
Nuclear science Symposium: Geant4 9.6, complemented by two correction patches.

The model implemented in Geant4 ``standard'' electromagnetic package 
\cite{emstandard} (also known as ``Sandia Table'') calculates
cross sections based on the analytical formula of Biggs and Lighthill, but it
reports using modified coefficients deriving from a fit to experimental data; 
nevertheless the reference cited in Geant4 9.6 Physics Reference Manual 
as the source of these modifications does not appear to be consistent.
Presumably, the modifications derive from \cite{grishin_1994}, which reports fits
to experimental data concerning noble gases, hydrogen, carbon, fluorine, oxygen and
silicon.
The energy of the emitted photoelectron is determined as the difference between
the energy of the interacting photon and the binding energy of the ionized shell
defined in the \emph{G4AtomicShells} class \cite{tns_binding}, 
and the photoelectron angle is calculated according to the Sauter-Gavrila distribution
for K shell \cite{sauter_1931,gavrila_1959}.

Geant4 low energy electromagnetic package \cite{lowe_chep,lowe_nss}
encompasses two implementations of the photoelectric effect, one identified as 
``Livermore'' \cite{lowe_e} and one reengineered from the 2008 version of
the Penelope code \cite{penelope2008}: both models calculate total and
partial cross sections based on EPDL97.
The so-called ``Livermore'' model provides three options of computing the
angular distribution of the emitted photoelectron: in the same direction as the
incident photon, based on Gavrila's distribution of the polar angle
\cite{gavrila_1959} for the K shell and the L$_1$ subshell, and
based on a double differential cross section derived from Gavrila's
\cite{gavrila_1959,gavrila_1961} calculations, which can also handle polarized
photons.

In addition, the Geant4 toolkit encompasses two models for the simulation of the
photoelectric effect concerning polarized photons: one for circularly polarised
photons in the ``polarisation'' package and one in the low energy
electromagnetic package, identified as ``Livermore polarised''.
Polarized photons are not considered in this study.

% -----------------------------------------------------------------------------------------

\section{Strategy of this study}

An extensive set of simulation models, which are representative of the variety
of theoretical and empirical approaches documented in the literature, have been evaluated to
identify the state-of-the-art of modeling photoionization in the context of
Monte Carlo particle transport.
%The physics models considered in this analysis involve the implementation of simple
%formulae in the simulation software or exploit available tabulations of complex
%theoretical calculations.

%The information summarized here concerns models implemented in Monte Carlo codes
%to describe the interaction of non-polarized photons; algorithms for the
%simulation of the photoelectric effect induced by polarized photons are not
%considered in this paper due the difficulty of retrieving experimental data in
%the literature for their validation.

%As it is highlighted in the introduction to this paper, physics models that are
%suitable for use in Monte Carlo particle transport

The models for the simulation of photoionization evaluated in this paper concern
total and partial cross sections: in particle transport, the former are
relevant to determine the occurrence of the photoionization process, while the latter
determine the the creation of a vacancy in a specific shell.

In addition, formulations of the photoelectron angular distribution have been evaluated.

All the models subject to study have been implemented in a consistent software
design, compatible with the Geant4 toolkit, which minimizes external
dependencies to ensure the unbiased appraisal of their intrinsic capabilities.

A wide set of experimental data of  has been collected
from the literature for this study; simulation models are validated through comparison
with these measurements.
The compatibility with experiment for each model, and the differences in
compatibility with experiment across the various models, are quantified by means
of statistical methods.

% -----------------------------------------------------------------------------------------

\subsection{Software environment}
\label{sec_sw}

All the physics models evaluated in this paper have been implemented in the same
software environment, which is compatible with Geant4; computational features
specific to the original physics algorithms have been preserved as much as possible. 
The uniform software configuration ensures an unbiased appraisal of the
intrinsic characteristics of the various physics models.
The correctness of implementation has been verified prior to the validation 
process to ensure that the software reproduces the physical features
of each model consistently.

The software adopts a policy-based class design \cite{alexandrescu}; this 
technique was first introduced in a general-purpose Monte Carlo system in
\cite{tns_dna}.

%Preliminary evaluations \cite{em_chep2009,em_nss2009} indicate that policy-based
%design contributes to achieve better computational performance than conventional
%inheritance in physics calculations for particle transport, thanks to compile-time
%binding.

Two policies have been defined for the simulation of photoionization,
corresponding to the calculation of total cross section and to the generation of
the final state; they conform to the prototype design described in
\cite{em_chep2009,em_nss2009}.
A photoionization process, derived from the \textit{G4VDiscreteProcess}
class of Geant4 kernel, acts as a host class for these policy classes.
All the simulation models implemented according to this policy-based class design
are compatible for use with Geant4, since Geant4 tracking handles 
all discrete processes polymorphically through the \textit{G4VDiscreteProcess} 
base class interface.

A single policy class calculates total cross sections for all the physics models
that exploit tabulations; alternative tabulations, corresponding to different
physics models, are managed through the file system.
Specific policy classes implement the analytical calculations of Biggs and Lighthill (accounting for the 
modifications adopted in Geant4 ''standard'' electromagnetic package) and of Ebel. 

Three photoelectron angular distribution models have been implemented:
they 
correspond to the Sauter-Gavrila formulation as in Geant4 ``standard'' 
electromagnetic package, to the Sauter-Gavrila formulation as in the 
\textit{G4PhotoElecricAngularGeneratorPolarized} class of Geant4 low
energy elecromagnetic package, and to a formulation based on corrected GEANT 3 code.

The software design adopted in this study ensures greater flexibility than the
design currently adopted in Geant4 electromagnetic package, since it allows
independent modeling and test of the various physics features of photoionization.
%Since policy classes are characterized by a single responsibility and have
%minimal dependencies on other parts of the software, the adopted programming paradigm
%facilitates the validation process.

% -----------------------------------------------------------------------------------------

\subsection{Experimental data}
\label{sec_exp}

%The validation of photon elastic scattering simulation models requires the
%availability of an extensive sample of experimental measurements, covering a
%wide range of energies, scattering angles and target elements.
Experimental data for the validation of the simulation models were collected
from a survey of the literature.
Only cross sections that were directly measured were considered in the validation process;
semi-empirical evaluations, derived from experimental measurements from 
which theoretical scattering cross section were subtracted to extract
photoelectric cross sections, were not considered.

The sample of experimental cross sections consists of more than 
5000 measurements: approximately 3700 total cross sections and 1400
shell cross sections, respectively.
Due to the limited page allocation typical of conference proceedings,
the extensive bibliographical references of the experimental data sample 
will be included in a forthcoming publication to be submitted to IEEE
Transactions on Nuclear Science.

%%Accuracies of edge positions are limited by chemical shifts and the detailed
%structure of the experimental material observed. Usually an accuracy of absolute
%energies below 1 - 3 eV is unattainable for this reason. At low energies (less
%than 200 - 500 eV) the occurrence of collective valence effects and dipole
%resonances can lead to much larger deviations (up to e.g., 50 eV or 10\%).
%
% 

% -----------------------------------------------------------------------------------------

\subsection{Data analysis method}
\label{sec_method}

The evaluation of the simulation models performed in this study has two
objectives: to validate them quantitatively, and to compare their relative capabilities.

The scope of the software validation process is defined according to the
guidelines of the pertinent IEEE Standard \cite{ieee_vv}.
For the problem domain considered in this paper, the validation process provides
evidence that the software models photoionization consistent with
experiment.

The analysis of cross sections is articulated over two stages: the
first one estimates the compatibility between the values calculated by
each simulation model and experimental data, while the second exploits the
results of the first stage to determine whether the various models exhibit any
significant differences in their compatibility with experiment.

The first stage encompasses a number of test cases, each one
corresponding to a photon energy and target element for which experimental data are available.
For each test case, cross sections calculated by the software are compared with
experimental measurements by means of goodness-of-fit tests; the null hypothesis  
is defined as the equivalence of the simulated and experimental data
distributions subject to comparison.

The goodness-of-fit analysis is based on the $\chi^2$ test \cite{bock} 
and utilizes the Statistical Toolkit \cite{gof1,gof2}.
The level of significance is 0.01.
%Among goodness-of-fit tests, this test has the peculiarity of taking into account
%experimental uncertainties explicitly; therefore the test statistic is sensitive
%to their correct appraisal.
The ``efficiency'' of a physics model is defined as the fraction of test cases
in which the $\chi^2$ test does not reject the null hypothesis at 0.01
level of significance.
%
%: it quantifies the capability of that simulation model to
%produce results statistically consistent with experiment over the whole set of
%test cases, which in physical terms means over the whole energy range of the 
%experimental data sample involved in the validation process.

The second stage of the statistical analysis quantifies the differences of the
simulation models in compatibility with experiment.
It consists of a categorical analysis based on contingency tables, which derive
from the results of the $\chi^2$ test: the outcome of this test is classified as
``fail'' or ``pass'', according respectively to whether the hypothesis of
compatibility of experimental and calculated data is rejected or not.
%The categorical analysis takes as a reference the simulation model exhibiting
%the largest efficiency at reproducing experimental data; the other models are
%compared to it, to determine whether they exhibit statistically significant
%differences of compatibility with measurements.

The null hypothesis in the analysis of a contingency table assumes the
equivalent compatibility with experiment of the models it compares.
Contingency tables are analyzed with Fisher's exact test \cite{fisher},
Barnard's exact test \cite{barnard} and
Pearson's $\chi^2$ test \cite{pearson} (the last one when appropriate).
The use of different tests  mitigates the risk of introducing
systematic effects, which could be due to the peculiar mathematical properties of
a single test.
 
The significance level for the rejection of the null hypothesis in the analysis
of contingency tables is 0.05, unless differently specified.

Due to the scarcity of experimental data and the unclear systematics in the measurements, 
a statistical analysis of photoelectron angular distribution would not be meaningful.
For this observable the comparison with experimental data is limited to qualitative appraisal.

% -----------------------------------------------------------------------------------------

\section{Results}
\label{sec_results}

Only a brief summary of the results of the validation process is reported here;
the full set of results will be documented in detail in a forthcoming journal
publication.

% -----------------------------------------------------------------------------------------

\subsection{Total Cross Sections}
\label{sec_cross}

\begin{figure}
\centerline{\includegraphics[angle=0,width=8.5cm]{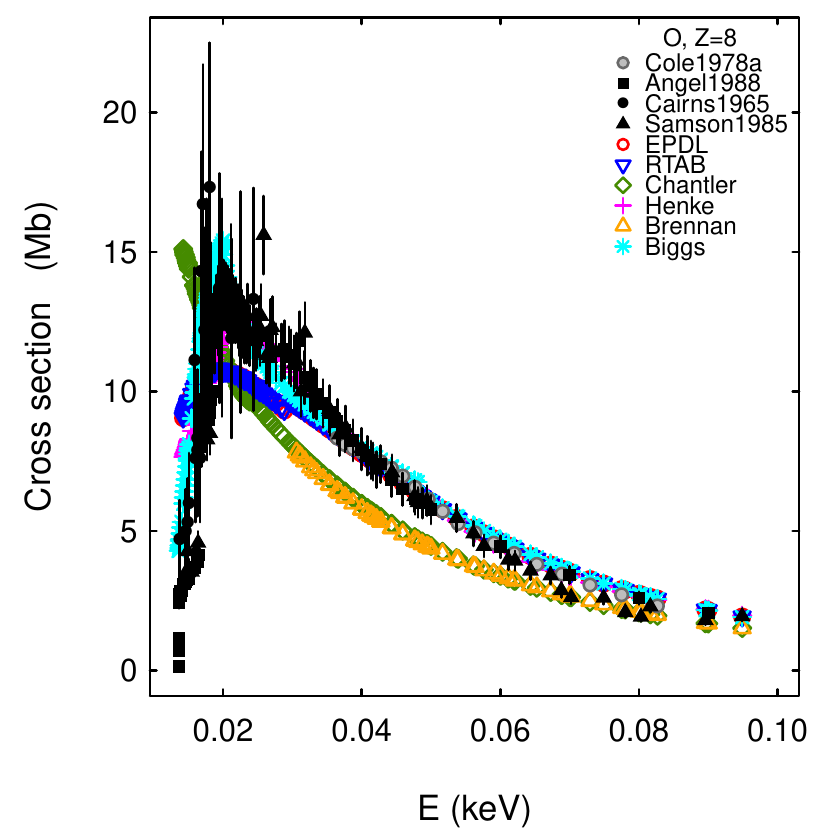}}
\caption{Total photoionization cross section for Z=8 as a function of photon energy.}
\label{fig_totO}
\end{figure}

\begin{figure}
\centerline{\includegraphics[angle=0,width=8.5cm]{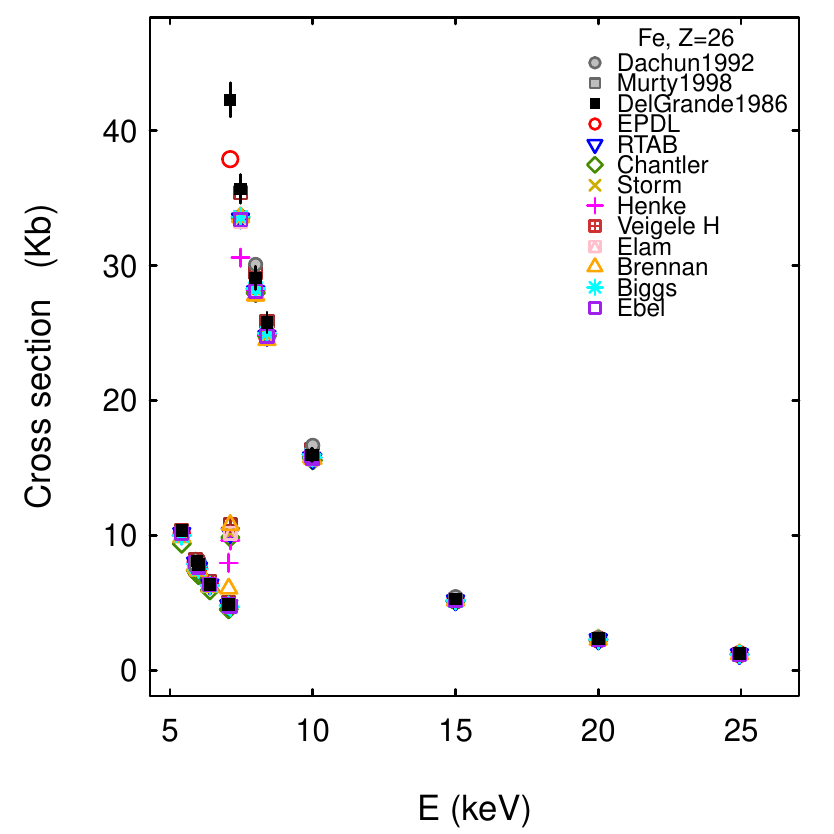}}
\caption{Total photoionization cross section for Z=26 as a function of photon energy.}
\label{fig_totFe}
\end{figure}

Figs. \ref{fig_totO} and \ref{fig_totFe} illustrate two examples of calculated and experimental total cross sections.

\begin{figure}
\centerline{\includegraphics[angle=0,width=8.5cm]{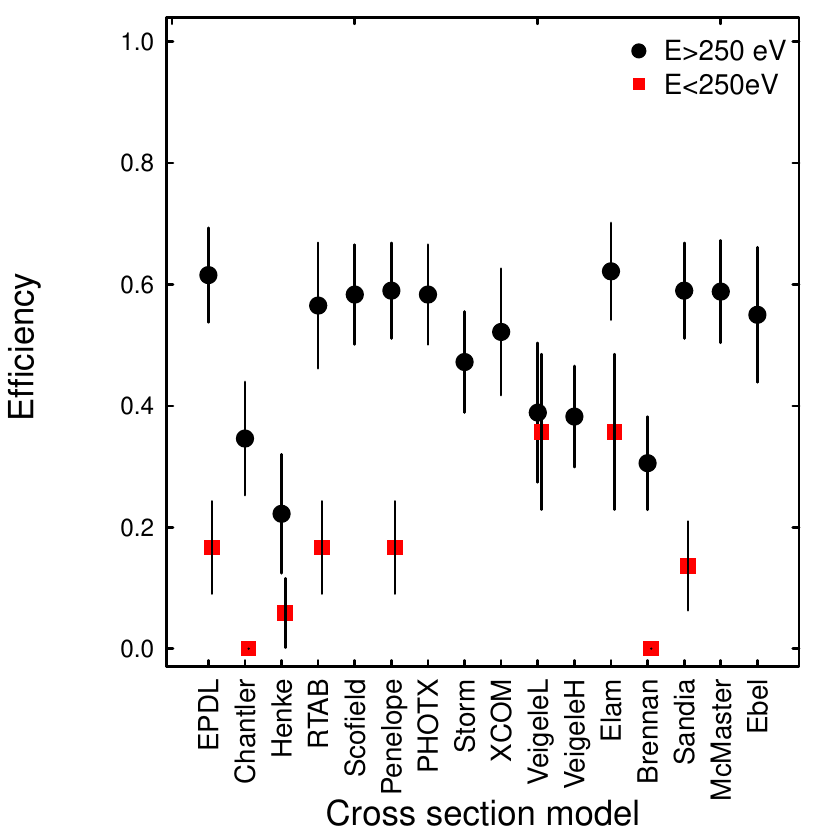}}
\caption{Efficiency of the total cross section calculation methods subject to test.}
\label{fig_efftot}
\end{figure}

The ``efficiency'' of the various total cross section calculation methods is documented in 
Fig. \ref{fig_efftot}: most methods exhibit similar compatibility with experiment for photon 
energies greater than 250 eV, while degraded accuracy is observed at lower energies.
Two cross section calculation methods exhibit lower compatibility with experiment in 
Fig. \ref{fig_efftot}; this qualitative observation is confirmed quantitatively by the 
results of the analysis of contingency tables reported in Table \ref{tab_conttot},
 where their compatibility with experimental data is
compared with that of EPDL: for both models all the tests applied to the associated contingency 
tables show that the hypothesis of equivalent performance with respect to EPDL is rejected with 0.05 significance.

\begin{table}[htbp]
  \centering
  \caption{P-values resulting from contingency tables comparing the compatibility with experiment of Chantler and EPDL total cross section calculations,
and of  Brennan and Cowan and EPDL calculations.}
    \begin{tabular}{lcc}
   \hline
    Test			& Chantler - EPDL	& Brennan and Cowan - EPDL\\
%			& EPDL 	&EPDL \\
    \hline
    Fisher		&  0.044 	& 0.011	  \\
    $\chi^2$ 	& 0.033	& 0.007		 \\
    Barnard		& 0.035	& 0.007	\\
\hline
    \end{tabular}%
  \label{tab_conttot}%
\end{table}%

% -----------------------------------------------------------------------------------------

\subsection{Shell Ionization Cross Sections}
\label{sec_shellcs}

\begin{figure}
\centerline{\includegraphics[angle=0,width=8.5cm]{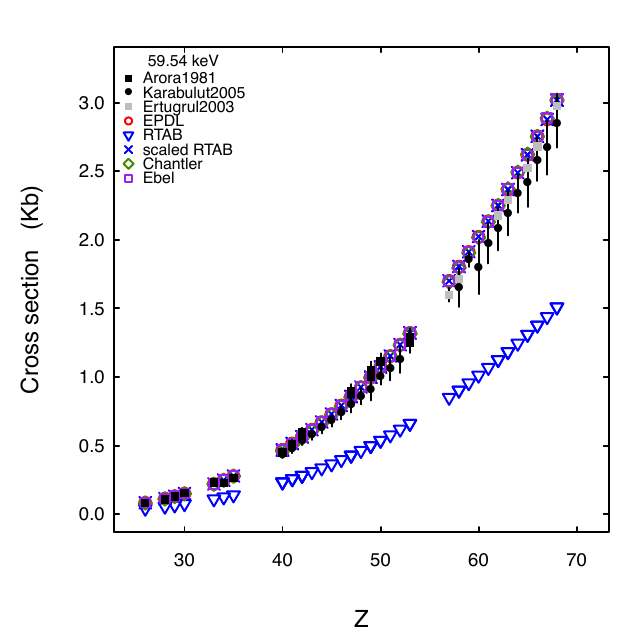}}
\caption{Cross section for photoionization of the K shell at 59.54 keV as a function of atomic number.}
\label{fig_K}
\end{figure}

\begin{figure}
\centerline{\includegraphics[angle=0,width=8.5cm]{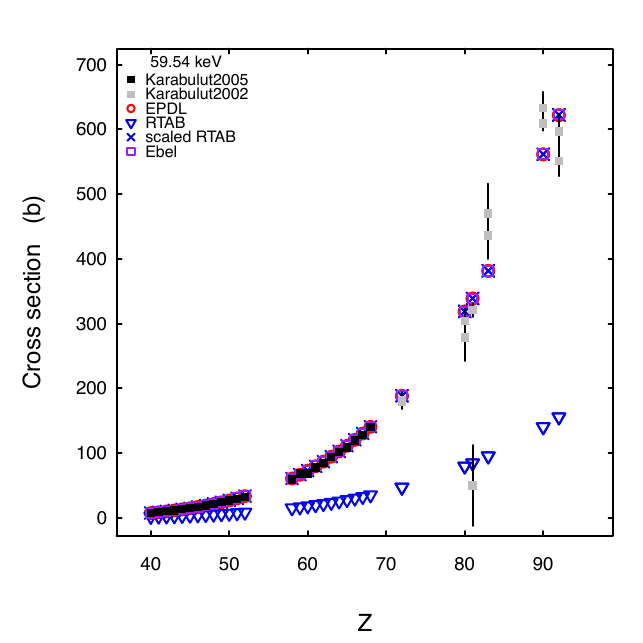}}
\caption{Cross section for photoionization of the L$_3$ subshell at 59.54 keV as a function of atomic number.}
\label{fig_L3}
\end{figure}

\begin{figure}
\centerline{\includegraphics[angle=0,width=8.5cm]{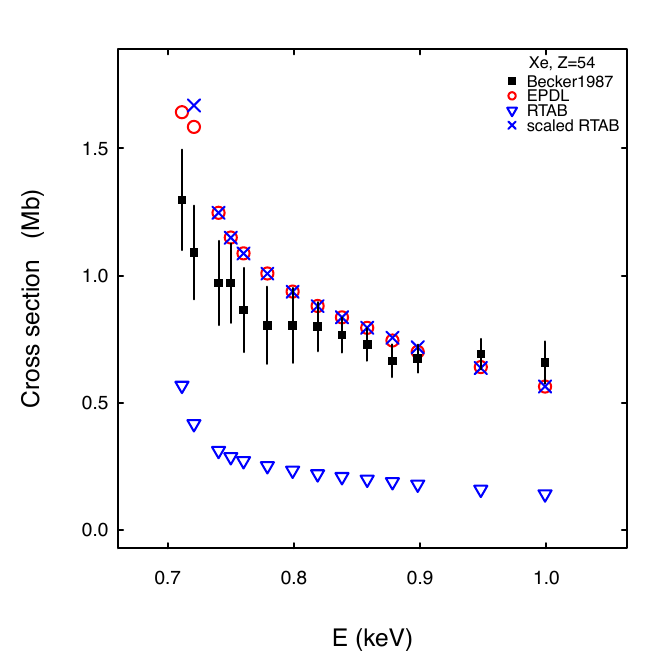}}
\caption{Cross section for photoionization of the M$_4$ subshell of xenon as a function of energy.}
\label{fig_M4}
\end{figure}

\begin{figure}
\centerline{\includegraphics[angle=0,width=8.5cm]{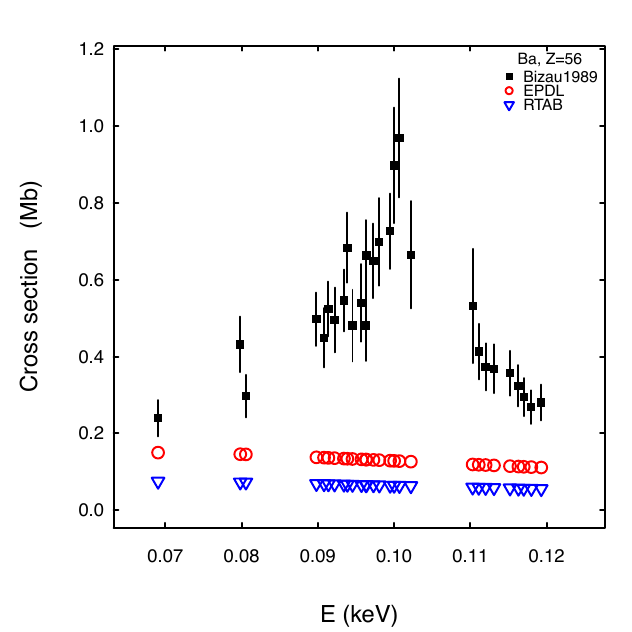}}
\caption{Cross section for photoionization of the O$_1$ subshell of barium as a function of energy.}
\label{fig_O1}
\end{figure}

Figs. \ref{fig_K} to \ref{fig_O1} illustrate some examples of calculated and experimental cross sections
for inner and outer shell photoionization.

A systematic discrepancy of RTAB shell cross sections with respect to
experimental data is observed, which hints to a missing multiplicative
factor in the tabulated values.
When RTAB cross sections are scaled by the presumed missing factor, 
they exhibit compatibility with experiment comparable to other calculation 
methods.

The p-values resulting from the $\chi^2$ test listed in Table \ref{tab_pshell}
show that, once RTAB values are rescaled, all calculation methods determine 
K and L shell cross sections that are compatible with experimental data with 
0.05 significance, with the exception of Ebel's parameterized model.
The cross sections for outer shells appear incompatible with experiment; 
nevertheless one should take into account that the experimental data samples 
available for the validation of outer shells are small, and often the data for a given 
test case originate from a single experimental source, which could be affected by 
systematic effects.

% Table generated by Excel2LaTeX from sheet 'Rexp-shell'
\begin{table}[htbp]
  \centering
  \caption{P-values of the $\chi^2$ test for compatibility of shell cross sections with experimental data }
    \begin{tabular}{lccccc}
    \hline
    \textbf{shell} & \textbf{EPDL} & \textbf{Chantler} & \textbf{RTAB} & \textbf{RTAB} & \textbf{Ebel} \\
& & & & (scaled) & \\
    \hline
    \textbf{K} & 0.209 & 0.350 & $<0.001$ & 0.315 & $<0.001$ \\
    \textbf{L$_1$} & 0.075 &       & $<0.001$ & 0.069 & 0.964 \\
    \textbf{L$_2$} & 0.339 &       & $<0.001$ & 0.299 & 0.154 \\
    \textbf{L$_3$} & 1     &       & $<0.001$ & 1     & 1 \\
    \textbf{M$_1$} & $<0.001$ &       & $<0.001$ & $<0.001$ &  \\
    \textbf{M$_4$} & 0.031 &       & $<0.001$ & $<0.001$ &  \\
    \textbf{M$_5$} & $<0.001$ &       & $<0.001$ & $<0.001$ &  \\
    \textbf{N$_1$} & $<0.001$ &       & $<0.001$ & $<0.001$ &  \\
    \textbf{N$_6$} & $<0.001$ &       & $<0.001$ & $<0.001$ & $<0.001$ \\
    \textbf{N$_7$} & $<0.001$ &       & $<0.001$ & $<0.001$ & $<0.001$ \\
    \textbf{O$_1$} & $<0.001$ &       & $<0.001$ & $<0.001$ & $<0.001$ \\
    \textbf{O$_2$} & $<0.001$ &       & $<0.001$ & $<0.001$ & $<0.001$ \\
    \textbf{O$_3$} & $<0.001$ &       & $<0.001$ & $<0.001$ & $<0.001$ \\
    \textbf{P$_1$} & $<0.001$ &       & $<0.001$ & $<0.001$ & $<0.001$ \\
    \hline
    \end{tabular}%
  \label{tab_pshell}%
\end{table}

% -----------------------------------------------------------------------------------------

\subsection{Photoelectron Angular Distribution}
\label{sec_angular}

\begin{figure}
\centerline{\includegraphics[angle=0,width=8.5cm]{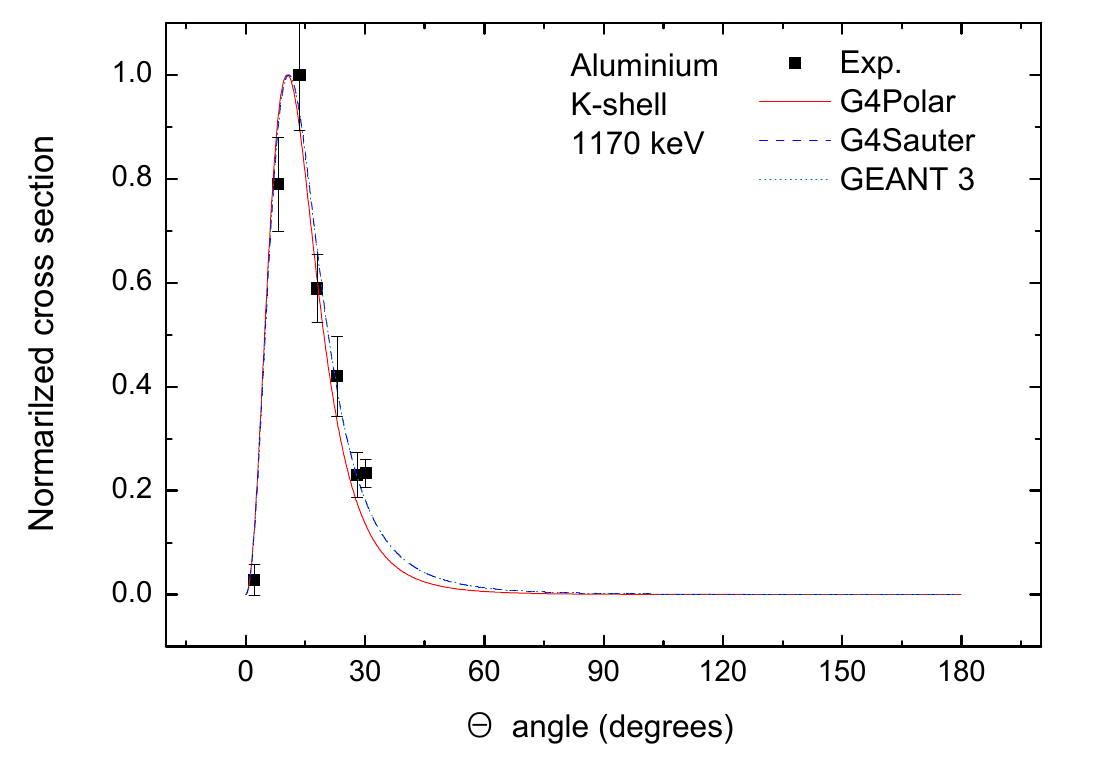}}
\caption{Photoelectron angular distribution for aluminium, K shell, at 1.17 MeV.}
\label{fig_angK}
\end{figure}

\begin{figure}
\centerline{\includegraphics[angle=0,width=8.5cm]{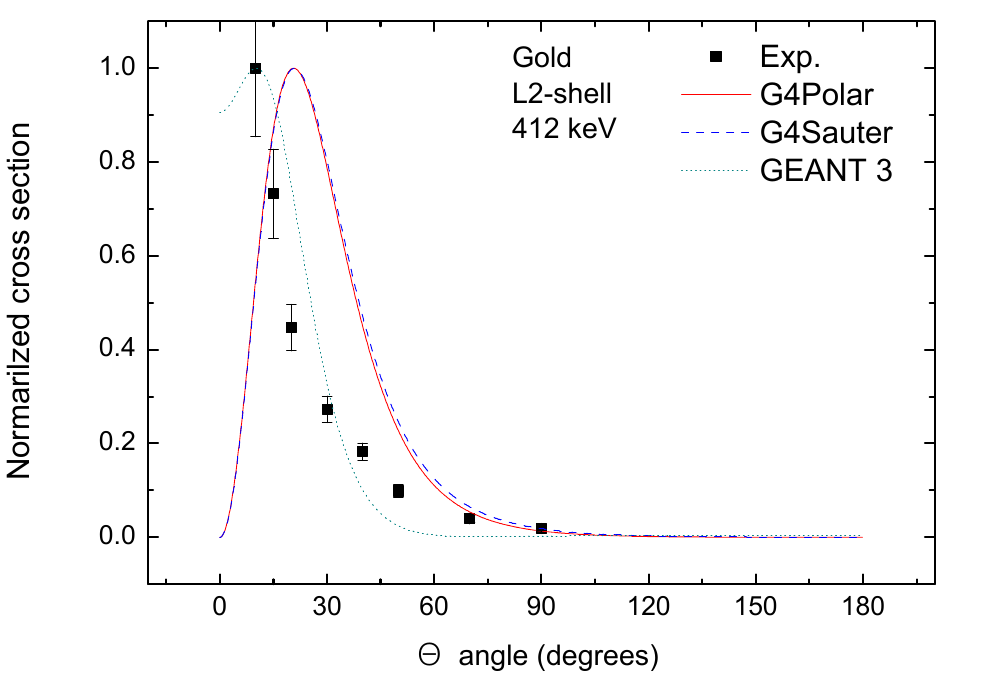}}
\caption{Photoelectron angular distribution for gold, L$_2$ subshell, at 412 keV.}
\label{fig_angL2}
\end{figure}

Two examples of angular distributions are shown in Figs. \ref{fig_angK} and \ref{fig_angL2}.
The limited experimental data sample does not allow a meaningful statistical analysis.

% -----------------------------------------------------------------------------------------

% -----------------------------------------------------------------------------------------

\section{Conclusion}

An extensive set of models for the simulation of photoionization has been
quantitatively evaluated regarding their accuracy at reproducing experimental
measurements.

All total cross section calculation methods are equivalent in terms of compatibility with
experimental data, with the exception of those calculated by Chantler and by Brennan
and Cowan. 
The fraction of test cases that are compatible with experiment drops at energies 
below 250 eV.

Inner shell cross section calculations are compatible with experimental data, with the exception 
of Ebel's parameterization for the K shell.
Outer shell photoionization cross sections are incompatible with experimental data; nevertheless 
the limited data sample hints not to draw any hasty conclusions about the
accuracy of the examined models.

Due to the scarce availability of experimental data and possible systematic effects in the
reported measurements, only a qualitative appraisal of photoelectron angular distribution
models is possible.
All Geant4 angular distribution models exhibit a similar behavior; the corrected GEANT 3
model appears in some cases different from the others and qualitatively competitive.

The full set of results will be documented in a forthcoming journal publication.

% --------------------------------------------------------------------------

% ------------------------------------------------------------------------------
\section*{Acknowledgment}
%The authors express their gratitude to CERN for support to the research
%described in this paper.

The CERN Library has provided helpful assistance
and essential reference material for this study.
%The authors thank ?? for proofreading the manuscript and valuable comments.

% ------------------------------------------------------------------------


\begin{thebibliography}{199}

%\bibitem{cesareo1992} 
%R. Cesareo, A. L. Hanson, G. E. Gigante, L. J. Pedraza, and S. Q. G. Mahtaboally,
%``Interaction of keV photons with matter and new applications'',
%\textit{Phys. Rep.}, vol. 213, no. 3, pp. 117-178, 1992.

% Reviews

\bibitem{pratt_1973}
R. H. Pratt, A. Ron, and H. K. Tseng, 
``Atomic photoeffect above 10 keV'',
\textit{Rev. Mod. Phys.}, vol. 45, 273-325, 1973.

\bibitem{pratt_1973_err}
R. H. Pratt, A. Ron, and H. K. Tseng, 
``Erratum: Atomic photoeffect above 10 keV'',
\textit{Rev. Mod. Phys.}, vol. 45, 663-664, 1973.

\bibitem{samson_1976}
J. A. R. Samson,
``Photoionization of atoms and molecules'',
\textit{Phys. Rep.}, vol. 28, no. 4, pp. 303-354, 1976.

\bibitem{kelly_1990}
H. P. Kelly,
``Review of our present understanding of the photoionization process for atoms'',
\textit{AIP Conf. Proc.}, vol. 215, pp. 292-311, 1990.

\bibitem{starace_2006}
A. F. Starace,
``Photoionization of atoms'',
in \textit{Atomic, Molecular and Optical Physics Handbook},Springer, Berlin, pp. 379-390, 2006.

%Books
\bibitem{amusia_1990}
M. Y. Amusia,
``Atomic Photoeffect'',
Plenum, New York, 1990.

\bibitem{berkowitz_2002}
J. Berkowitz,
``Atomic and Molecular Photoabsorption'',
Academic Press, London, 2002.

% Photons in MC in general

%\bibitem{hubbell_2006}
%J. H. Hubbell,
%``Review and history of photon cross section calculations'',
%\textit{Phys. Med. Biol.}, vol. 51, pp. R245-R262, 2006.


% EM validation

\bibitem{tns_nist}
K. Amako et al., 
``Comparison of Geant4 electromagnetic physics models against the NIST 
reference data'', 
\emph{IEEE Trans. Nucl. Sci.}, vol. 52, no. 4, pp. 910-918, 2005.

\bibitem{chica_2009}
U. Chica, M. Anguiano, and A. M. Lallena,
``Benchmark of PENELOPE for low and medium energy X-rays'',
\textit{Phys. Med.}, vol. 25, no. 2, pp. 51–57, 2009.
    

\bibitem{trucano_what}
T. G. Trucano, L. P. Swiler, T. Igusa, W. L. Oberkampf, and M. Pilch,
``Calibration, validation, and sensitivity analysis: What’s what",
\textit{Reliab. Eng. Syst. Safety}, vol. 91, no. 10-11, pp. 1331-1357, 2006.

\bibitem{cirrone2010}
G. A. P. Cirrone, G. Cuttone, F. Di Rosa, L. Pandola, F. Romano, and Q. Zhang,
``Validation of the Geant4 electromagnetic photon cross-sections for elements and compounds'',
\textit{Nucl. Instr. Meth. A}, vol. 618, pp. 315-322, 2010.

% Basic Geant4 references:
\bibitem{g4nim} 
S.~Agostinelli et al., 
``Geant4 - a simulation toolkit''
\textit{Nucl. Instrum. Meth. A}, vol. 506, no. 3, pp. 250-303, 2003.

\bibitem{g4tns}
J.~Allison et al., 
``Geant4 Developments and Applications'' 
\textit{IEEE Trans. Nucl. Sci.}, vol. 53, no. 1, pp. 270-278, 2006.

% Atomic relaxation

\bibitem{tns_relax}
S. Guatelli, A. Mantero, B. Mascialino, P. Nieminen, and M. G. Pia, 
``Geant4 Atomic Relaxation'', 
\emph{IEEE Trans. Nucl. Sci.}, vol. 54, no. 3, pp. 585-593, 2007.

\bibitem{tns_relax_nist}
S. Guatelli et al., 
``Validation of Geant4 Atomic Relaxation against the NIST Physical Reference Data'', 
\textit{IEEE Trans. Nucl. Sci.}, vol. 54, no. 3, pp. 594-603, 2007.

\bibitem{tns_relax_prob}
M. G. Pia, P. Saracco, and M. Sudhakar,
``Validation of K and L Shell Radiative Transition Probability Calculations'',
\textit{IEEE Trans. Nucl. Sci.}, vol. 56, no. 6, pp. 3650-3661,  2009.

\bibitem{tns_binding}
M. G. Pia et al.,
``Evaluation of atomic electron binding energies for Monte Carlo particle transport'',
\textit{IEEE Trans. Nucl. Sci.}, vol. 58, no. 6, pp. 3246-3268, 2011.

% Compounds

%\bibitem{egs_compound1}
%H. Hirayama and Y. Namito,
%``Implementation of a General Treatment of Photoelectric-Related Phenomena for Compounds or Mixtures in EGS4'',
%KEK Internal 2000-3 Report, Tsukuba, Japan, 2000.
%
%\bibitem{egs_compound2}
%H. Hirayama and Y. Namito,
%``Implementation of a General Treatment of Photoelectric-Related Phenomena for Compounds or Mixtures in EGS4 (Revised Version)'',
%KEK Internal 2004-6 Report, Tsukuba, Japan, 2004.

% Cross section calculations

\bibitem{pratt_1960}
R. H. Pratt, 
``Atomic photoelectric effect at high energies'', 
\textit{Phys. Rev.}, vol. 117, pp. 1017-1028, 1960.

\bibitem{pratt_1964}
R. H. Pratt,
``K-Shell Photoelectric Cross Sections from 200 keV to 2 MeV'',
\textit{Phys. Rev.}, vol. 134, pp. A898-A915, 1964.

\bibitem{rakavy_1967}
G. Rakavy and A. Ron,
``Atomic Photoeffect in the Range E$_{\gamma}$=1-2000~keV'',
\textit{Phys. Rev.}, vol. 159, pp. 50-56, 1960.

\bibitem{schmickley_1967}
R. D. Schmickley and R. Pratt, 
``K-, L-, and M-shell atomic photoeffect for screened-potential models'',
\textit{Phys. Rev}., vol. 164, pp. 104-116, 1967.

\bibitem{scofield_1973}
J. H. Scofield, 
``Theoretical photoionization cross sections from 1 to 1500 keV'',
Report UCRL-51326, Lawrence Livermore Laboratory, 1973.

\bibitem{chantler_1995}
C. T. Chantler,
``Theoretical form factor, attenuation and scattering tabulation for Z=1–92 from
E=1–10 eV to E=0.4–1.0 MeV'',
\textit{J. Phys. Chem. Ref. Data}, vol. 24, no. 1, pp. 71–643, 1995.

\bibitem{chantler_2000}
C. T. Chantler,
``Detailed tabulation of atomic form factors, photoelectric absorption and
scattering cross section, and mass attenuation coefficients in the vicinity of
absorption edges in the soft X-ray (Z= 30-36, Z= 60-89, E= 0.1 keV-10 keV),
addressing convergence issues of earlier work'',
\textit{J. Phys. Chem. Ref. Data}, vol. 29, no. 4, pp. 597-1048, 2000.

\bibitem{hubbell_1969}
J. H. Hubbell, 
``Photon Cross Sections, Attenuation Coefficients, and Energy Absorption Coefficients from 10~keV to 100~GeV'', 
Report NSRDS-NBS 29, National Bureau of Standards, Washington, DC (USA), 1969.

\bibitem{biggs3}
F. Biggs and R. Lighthill,
``Analytical Approximations for X-RayCrossSections III''
Sandia National Laboratories Report SAND87-0070, Albuquerque, 1988.

% Ebel

\bibitem{ebel_2003}
 H. Ebel, R. Svagera, M. F. Ebel, A. Shaltout, and  J. H. Hubbell,
``Numerical description of photoelectric absorption coefficients for fundamental parameter programs'',
\textit{X-Ray Spectrom.}, vol. 32, no. 6, pp. 442–451, 2003.


% Angular distribution

\bibitem{fischer_1931}
F. Fischer,
``Beitr{\"a}ge zur Theorie der Absorption von R{\"o}ntgenstrahlung'',
\textit{Ann. Phys.}, vol. 400, no. 7, pp. 821-850, 1931.

\bibitem{sauter_1931}
F. Sauter, ``{\"U}ber den atomaren Photoeffekt in der K-Schale nach der relativistischen Wellenmechanik Diracs'',
\textit{Ann. Phys.}, vol. 403, no. 4, pp. 454-488, 1931.

\bibitem{sauter_1931a}
F. Sauter,
``{\"U}ber den atomaren Photoeffekt bei gro{\ss}er H{\"a}rte der anregenden Strahlung'',
\textit{Ann. Phys.}, vol. 401, no. 2, pp. 217-248, 1931.

\bibitem{gavrila_1959}
M. Gavrila,
``Relativistic k-shell photoeffect'',
\textit{Phys. Rev.}, vol. 113, pp. 514-536, 1959.

\bibitem{nagel_1963}
B. Nagel,
``Angular distribution and polarization of K-shell photoelectrons in the high energy limit'',
\textit{ Ark. Fys.}, vol. 24, pp. 151-159, 1963.

\bibitem{gavrila_1961}
M. Gavrila, 
``Relativistic l-shell photoeffect'',
\textit{Phys. Rev.}, vol. 124, pp. 1132-1141, 1961.


% Other compilations

% Brennan-Cowan
\bibitem{brennan_cowan_1992}
S. Brennan and P. L. Cowan,
``A suite of programs for calculating xray absorption, reflection, and 
diffraction performance for a variety of materials at arbitrary wavelengths'',
\textit{Rev. Sci. Instrum.}, vol., 63, pp. 850-853, 1992.

\bibitem{elam_2002}
W. Elam, B. Ravel, and J. Sieber, 
``A new atomic database for X-ray spectroscopic calculations'',
\textit{Radiat. Phys. Chem.}, vol. 63, pp. 121-128, 2002.

\bibitem{epdl97}	
D. Cullen et al., 
``EPDL97, the Evaluated Photon Data Library'', 
Lawrence Livermore National Laboratory Report UCRL-50400, Vol. 6, Rev. 5, 1997.

% Henke
\bibitem{henke_1982}
B. L. Henke, P. Lee, T. Tanaka, R. L. Shimabukuro, and B. Fujikawa, 
``Low-energy X-ray interaction coefficients: Photoabsorption, scattering, and reflection: E= 100–2000 eV Z= 1–94'',
\textit{Atom. Data Nucl. Data Tables}, vol. 27, pp. 1-144, 1982.

\bibitem{henke_1993}
 B. L. Henke, E. M. Gullikson, and J. C. Davis
``X-Ray Interactions: Photoabsorption, Scattering, Transmission, and Reflection at E = 50-30000 eV, Z = 1-92'',
\textit{Atom. Data Nucl. Data Tables}, vol. 54, no. 2, pp. 181–342, 1993.

% McMaster

\bibitem{mcmaster_1969}
W. H. McMaster, N. Kerr Del Grande, J. H. Mallet and J. H. Hubbell, 
``Compilation of X-ray cross sections'',  Section II Revision 1,
Lawrence Livermore National Laboratory Report UCRL-50174, 1969.

%\bibitem{mcmaster_news}
%W. H. McMaster, N. Kerr Del Grande, J. H. Mallet and J. H. Hubbell, 
%``Compilation of x-ray cross sections UCRL-50174, sections I, II revision 1, III, IV'',
%\textit{Atom. Data Nucl. Data Tables}, vol. 8, no. 4–6, pp. 443–444, 1970.

\bibitem{shaltout_2006}
A. Shaltout, H. Ebel, and R. Svagera, 
``Update of photoelectric absorption coefficients in the tables of McMaster'',
\textit{X-Ray Spectrom.}, vol. 35, pp. 52-56, 2006.

\bibitem{photx}
D. K. Trubey,  M. J. Berger, and J. H. Hubbell, 
``Photon cross sections for ENDF/B-VI'',
Oak Ridge National Lab. Report CONF-890408-4, 1989.

\bibitem{rtab}
L. Kissel,
``RTAB: the Rayleigh scattering database'',
\textit{Radiat. Phys. Chem.}, vol. 59, pp. 185-200, 2000.

% Storm-Israel

\bibitem{storm_israel_photon}
E. Storm and H. I. Israel,
``Photon cross sections from 1 keV to 100 MeV for elements Z=1 to Z=100'',
\textit{Atom. Data Nucl. Data Tables}, vol. 7, pp. 565-681, 1970.

\bibitem{veigele_1973}
W. M. J. Veigele,
``Photon cross sections from 0.1 keV to 1 MeV for elements Z = 1 to Z= 94'',
\textit{Atom. Data Nucl. Data Tables}, vol. 5, no. 1, pp. 51-111, 1973.

\bibitem{xcom}
M. J. Berger et al.,
 ``XCOM: Photon Cross section Database (version 1.5)'', 
National Institute of Standards and Technology, Gaithersburg, MD, 2010.
[Online] Available: http://physics.nist.gov/xcom.




%\bibitem{saloman_1988}
%E. B. Saloman, J. H. Hubbell and J. H. Scofield, 
%``X-Ray Attenuation cross sections for Energies 100 eV to 100 keV and Elements Z=1 to Z=92'',
% \textit{Atom. Data Nucl. Data Tables}, vol. 38, pp. 1-197, 1988.
%
%\bibitem{saloman_1986}
%E. B. Saloman and J. H. Hubbell,
%``X-ray attenuation coefficients (total cross sections): Comparison of the
%experimental data base with the recommended values of Henke and the theoretical
%values of Scofield for energies between 0.1-100~keV'',
%Report NBSIR-86-3431, National Bureau of Standards, Washington, DC (USA), 1986.

% Data libraries


% MC codes

% EGS

\bibitem{egs4}
W. R. Nelson, H. Hirayama, and D. W. O. Rogers, 
``The EGS4 Code System'',
SLAC-265 Report, Stanford, CA, 1985. 

\bibitem{sakamoto}
Y. Sakamoto,
``Photon cross section data PHOTX for PEGS4 code'',
 in \textit{Proc.Third EGS4 User’s Meeting in Japan}, KEK Proceedings 93-15, pp. 77–82, Japan, 1993.

\bibitem{egs4_angular}
A. F. Bielajew and D. W. O. Rogers,
``Photoelectron angular distribution in the EGS4 code system'',
Report PIRS-0058, National Research Council of Canada, 1986.

\bibitem{egs5}
H. Hirayama, Y. Namito, A. F. Bielajew, S. J. Wilderman, and W. R.
Nelson, 
``The EGS5 Code System'', SLAC-R-730 Report, Stanford, CA, 2006.

\bibitem{egsnrc}
I. Kawrakow ,E. Mainegra-Hing, D.W.O. Rogers, F. Tessier and B.R.B. Walters, 
``The EGSnrc Code System: Monte Carlo
Simulation of Electron and Photon Transport
NRCC PIRS-701, 5th printing, 2010. 


% ETRAN
\bibitem{etran}
S. M. Seltzer,
``Electron-Photon	Monte	 Carlo	Calculations: The ETRAN Code'',
\textit{Appl.	Radiat. Isot.}, vol. 42, no. 10, pp. 917-941, 1991.

% FLUKA
\bibitem{fluka1}
G. Battistoni et al.,
``The FLUKA code: description and benchmarking",
\textit{AIP Conf. Proc.}, vol. 896, pp. 31-49, 2007.
%A.~Fass{\`o} et al., ``The physics models of FLUKA: status and recent
%developments'', in {\sl Proc.\ Computing in High Energy and Nuclear
%Physics 2003 Conference (CHEP 2003)}, La Jolla, CA, USA, paper MOMT05.

\bibitem{fluka2}
A.~Ferrari et al., 
``Fluka: a multi-particle transport code'', 
Report CERN-2005-010, INFN/TC-05/11, SLAC-R-773, Geneva, Oct. 2005.

% ITS
\bibitem{its5}
B. C. Franke, R. P. Kensek and T. W. Laub,
"ITS5 theory manual", rev. 1.2,
Sandia Natl. Lab. Report SAND2004-4782, Albuquerque, 2004.

% MCNP

\bibitem{mcnp5}
X-5 Monte Carlo Team, 
``MCNP -- A General Monte Carlo N-Particle Transport Code, Version 5'', 
Los Alamos National Laboratory Report LA-UR-03-1987,  2003, revised 2008.

\bibitem{mcnpx27e}
D. B. Pelowitz et al.,
``MCNPX 2.7.E Extensions'',
Los Alamos National Laboratory Report LA-UR-11-01502,  2011.

%\bibitem{mcnpx}
%J. S.~Hendricks et al., 
%``MCNPX, Version 2.6.0'', 
%Los Alamos National Laboratory Report LA-UR-06-7991,  2008.

\bibitem{epdl89}
D. E. Cullen et al.,
``Tables and Graphs of Photon Interaction Cross Sections from 10 eV to 100 GeV
Derived from the LLNL Evaluated Photon Data Library (EPDL)'',
Lawrence Livermore National Laboratory Report UCRL-50400, Vol. 6, Rev. 4, 1989.

\bibitem{ENDFB-IV}
D. Garber Ed. ,
``ENDF/B Summary Documentation'', 
BNL-17541, 1975.


% Penelope

%\bibitem{penelope}
%J.~Baro, J.~Sempau, J. M.~Fern\'andez-Varea, and F.~Salvat,
%``PENELOPE, an algorithm for Monte Carlo simulation of the
%penetration and energy loss of electrons and positrons in
%matter'', 
%\emph{Nucl. Instrum. Meth. B}, vol. 100, no. 1, pp. 31-46, 1995.

\bibitem{sempau_1997}
J. Sempau, E. Acosta, J. Baro, J. M. Fernandez-Varea, and . Salvat,
``￼An algorithm for Monte Carlo simulation of coupled electron-photon transport'',
\emph{Nucl. Instrum. Meth. B}, vol. 132, pp. 377-390, 1997.

%\bibitem{penelope2001}
%F. Salvat, J.M. Fernandez-Varea, E. Acosta, and J. Sempau,
%``Penelope - A code system for Monte Carlo simulation of electron and
%photon transport'', 
%Proc. Workshop NEA, 2001.

\bibitem{penelope2008}
F. Salvat, J.M. Fernandez-Varea, and J. Sempau,
``Penelope - A code system for Monte Carlo simulation of electron and
photon transport'', 
Proc. Workshop NEA 6416, 2008.

\bibitem{penelope2011}
F. Salvat, J.M. Fernandez-Varea, and J. Sempau,
``Penelope-2011 - A code system for Monte Carlo simulation of electron and
photon transport'', 
Proc. Workhop NEA/NSC/DOC(2011)5, 2011.

%\bibitem{baro1994a}
%J. Bar\'o, M. Roteta, J. M. Fernandez-Varea, and F. Salvat,
%``Analytical cross sections for Monte Carlo simulation of photon transport'',
%\textit{Radiat. Phys. Chem.}, vol. 44, no. 5, pp. 531-552, 1994.

% GEANT 3
\bibitem{geant3}
"GEANT Detector Description and Simulation Tool'', 
CERN Program Library Long Writeup W5013, 1995.

%\bibitem{egs3}
%R. L. Ford and W. R. Nelson,
%``The EGS code system: computer programs for the Monte Carlo
%simulation of electromagnetic cascade showers (version 3)'',
%SLAC-210 Report, Stanford, CA, 1978. 


% Geant4 stuff

\bibitem{emstandard}
H. Burkhardt et al., 
``Geant4 Standard Electromagnetic Package'',
in \textit{Proc. 2005 Conf. on Monte Carlo Method: Versatility Unbounded
in a Dynamic Computing World}, Am. Nucl. Soc., USA, 2005.

\bibitem{grishin_1994}
V. M. Grishin, A. P. Kostin, S. K. Kotelnikov and D. G. Streblechenko,
Parametrization of the photoabsorption cross section at low energies'',
\textit{Bull. Lebedev Inst.}, no. 6, pp. 1-6, 1994. 


\bibitem{lowe_chep}
S. Chauvie, G. Depaola, V. Ivanchenko, F. Longo, P. Nieminen and M. G. Pia,
``Geant4 Low Energy Electromagnetic Physics'',
in \textit{Proc. Computing in High Energy and Nuclear Physics}, 
Beijing, China, pp. 337-340, 2001.

\bibitem{lowe_nss}
S. Chauvie et al., ``Geant4 Low Energy Electromagnetic Physics'',
in \textit{2004 IEEE Nucl. Sci. Symp. Conf. Rec.}, pp. 1881-1885, 2004.

\bibitem{lowe_e} 
J. Apostolakis, S. Giani, M. Maire, P. Nieminen, M.G. Pia, L. Urban,
``Geant4 low energy electromagnetic models for electrons and photons''
\textit{INFN/AE-99/18}, Frascati, 1999. 

%\bibitem{penelope2001}
%F. Salvat, J.M. Fernandez-Varea, E. Acosta, and J. Sempau,
%``Penelope - A code system for Monte Carlo simulation of electron and
%photon transport'', 
%Proc. Workshop NEA, 2001.

 %---- Total cross sections



% ---- Software
\bibitem{alexandrescu}
A. Alexandrescu, 
``Modern C++ Design'', Ed.: Addison-Wesley, 2001.

\bibitem{tns_dna}
S. Chauvie et al., 
"Geant4 physics processes for microdosimetry simulation: design foundation and
implementation of the first set of models",
\textit{IEEE Trans. Nucl. Sci.,} vol. 54, no. 6, pp. 2619-2628, 2007.

\bibitem{em_nss2009}
M. Augelli et al.,
"Research in Geant4 electromagnetic physics design, and its effects on
computational performance and quality assurance",
in \textit{2009 IEEE Nucl. Sci. Symp. Conf. Rec.}, pp. 177-180, 2009.

\bibitem{em_chep2009}
M. G. Pia et al.,
"Design and performance evaluations of generic programming techniques in a R\&D
prototype of Geant4 physics",
\textit{J. Phys.: Conf. Ser.}, vol. 219, pp. 042019, 2010.



% ---- Verification and Validation

\bibitem{ieee_vv}
IEEE Computer Society,
``IEEE Standard for Software Verification and Validation'', 
IEEE Std 1012-2004, Jun. 2005.

%\bibitem{iso12207}
%ISO/IEC, ``International Standard, Information Technology Software Life
%Cycle Process, IS0 12207'', 2008-2011.


% ---- Statistics
\bibitem{bock}
R. K. Bock and W. Krischer,
``The Data Analysis BriefBook '',
Ed. Springer, Berlin, 1998. 

\bibitem{gof1}
G. A. P. Cirrone et al., 
``A Goodness-of-Fit Statistical Toolkit'', 
\emph{IEEE Trans. Nucl. Sci.}, vol. 51, no. 5, pp. 2056-2063, 2004.

\bibitem{gof2}
B. Mascialino, A. Pfeiffer, M. G. Pia, A. Ribon, and P. Viarengo, 
``New developments of the Goodness-of-Fit Statistical Toolkit'', 
\emph{IEEE Trans. Nucl. Sci.}, vol. 53, no. 6, pp. 3834-3841,  2006.


\bibitem{fisher}
R. A. Fisher,
``On the interpretation of  $\chi^2$ from contingency tables, 
and the calculation of P'',
\textit{J. Royal Stat. Soc.}, vol. 85, no. 1, pp. 87-94, 1922. 

\bibitem{barnard}
 G. A. Barnard,
`` Significance tests for 2 × 2 tables'',
\textit{Biometrika}, vol. 34, pp. 123-138, 1947.

\bibitem{pearson}
K. Pearson,
``On the $\chi^2$ test of Goodness of Fit'',
\emph{Biometrika}, vol. 14, no. 1-2, pp. 186-191, 1922.


\end{thebibliography}
\end{document}